# Roughness corrections to the Casimir force: The importance of local surface slope

P.J. van Zwol, G. Palasantzas[*] and J. Th. M. De Hosson

Department of Applied Physics, Netherlands Institute for Metals Research and Zernike Institute for Advanced Materials, University of Groningen, Nijenborgh 4, 9747 AG Groningen, the Netherlands.

**Abstract**

This paper concentrates on a study where finite conductivity corrections are included in the theoretical description of the effects of roughness on the Casimir force. The roughness data were taken from gold films evaporated onto Silicon and polysterene spheres. We conclude that for a detailed comparison with experimental data, i.e. at the level of at least 5 % at short separations below 200 nm, the lateral dimensions of roughness for real films should be included in the theoretical considerations. Moreover, if the RMS roughness is considerable, high local surface slopes are shown to have a significant effect on the Casimir force.

Pacs numbers: 78.68.+m, 03.70.+k, 85.85.+j, 12.20.Fv

[*]Author to whom correspondence should be addressed: g.palasantzas@rug.nl

When the proximity between material objects ranges between nanometers up to a few micrometers, a regime is entered in which forces become operative that are *quantum mechanical* in nature, namely *van der Waals* and *Casimir forces* [1]. Because of its relatively short range, the Casimir force [1] is now starting to take on technological importance in the design and operation of micro/nanoelectromechanical systems (MEMS/NEMS), e.g., micro/nano switches, nanoscale tweezers or actuators [2-9]. High accuracy measurements by Lamoreaux with the use of torsion pendulum [7] initiated detailed investigations of the Casimir force. It was also measured accurately by other groups in the plate-sphere setup with the Atomic Force Microscope (AFM), and micro oscillator devices [8, 9, 10]. Other geometries (crossed cylinders [11], and parallel plates [12]) were also investigated.

For most of the measurements, the Proximity Force Approximation (PFA) was used perturbatively up to fourth order to calculate the roughness effect on the Casimir force [7-12]. However, the Casimir force is not additive, and both PFA and additive methods use only the RMS roughness to predict its influence [2]. While this is the most important factor, any lateral information of rough films has been ignored [13]. Numerical approaches today are rather limited to simple systems making them unsuitable for predicting roughness effects of real systems [14]. Recently a model was developed to incorporate roughness effects into scattering theory [15]. Due to the complexity of the calculations only the second order corrections were presented, showing, however, significant deviations from the PFA.

Evaporated metallic films, which are used to coat substrates for the force measurements, show in many cases the so-called self-affine random roughness [16]. The importance of self affine scaling and its relation to the Casimir force has been emphasized in [17]. However, finite conductivity contributions were ignored, and only analytic solutions in some limited cases were given. Here, we performed a study where finite conductivity corrections were taken into account using experimental optical data. The range was extended by fitting a Drude model into the infrared regime [2]. The roughness data were taken from gold (Au) films evaporated onto Si and polysterene spheres  The discussion will focus on the effect of self affine roughness within scattering theory in comparison to PFA results, with emphasis on the local surface slope.

Within Lifshitz theory the Casimir energy between real parallel flat mirrors with area A, separated a distance $L$, with reflection coefficient r($\Phi$) and $\Phi$ the imaginary frequency of the electromagnetic wave, is given by

$$E_{ppflat} = -\hbar A \sum_P \int \frac{d^2k}{4\pi^2} \int_0^\infty \frac{d\Phi}{2\pi} \ln[1 - r^p(k,\Phi)^2 e^{-2\kappa L}]. \qquad (1)$$

The integral in Eq. (1) is over all field modes of the wave vector k and $\Phi$. The index p denotes the transverse electric and magnetic (TE and TM) modes. $A$ is the average flat surface area. Roughness corrections to the Casimir energy within the scattering formalism [9] are formulated in terms of a roughness response function G(k) and the roughness power spectrum σ(k): $\delta E_{pp} = \int [d^2k/4\pi^2] G(k)\sigma(k)$ where G(k) is derived in [13, 15] yielding for the total energy $E_{pp,rough} = E_{ppflat} + \delta E_{pp}$ [13, 15]. The theory is valid under the following assumptions. First the lateral dimensions of the roughness must be much smaller than the system size, i.e. plate or sphere, which is usually the case. Second, the RMS roughness $w$ must be small compared to the separation distance $L$ ($w<<L$), and third, lateral roughness dimensions must be much larger than the vertical dimensions, or conversely the local surface slope of a film must be small ($\rho_{rms}<<1$) [7,9]. For force measurements by AFM, a sphere plate geometry is often used to avoid plate alignment problems [7-10]. In this case, the Casimir force is given by $F_C = (2\pi R/A)E_{PP}$.

A wide variety of surfaces exhibit the so-called self-affine roughness [16], which is characterized for isotropic surfaces by the RMS roughness amplitude $w = <[h(r)]^2>^{1/2}$ ($<h>=0$), the lateral correlation length $\xi$ (indicating the lateral feature size), and the roughness exponent $0<H<1$. Small values of $H\sim0$ correspond to jagged surfaces, while large values $H\sim1$ to a smooth hill – valley morphology [16]. For self-affine roughness the spectrum $\sigma(k)$ scales as $\sigma(k) \propto k^{-2-2H}$ if $k\xi>>1$, and $\sigma(k) \propto const$ if $k\xi<<1$ [16]. This scaling is satisfied by the analytic model [16, 18] $\sigma(k) = (AHw^2\xi^2)/(1+k^2\xi^2)^{1+H}$ with $A = 2/[1-(1+k_c^2\xi^2)^{-H}]$ and $k_c$ a lower roughness cutoff ($\sim1$ nm$^{-1}$). The local surface slope $\rho_{rms} = <(\nabla h)^2>^{1/2}$ is given in this case by the analytical form $\rho_{rms} = (w/\xi)\{(AH/2)([1+k_c^2\xi^2]^{1-H}-1)/(1-H)-1/2\}^{1/2}$. The parameters $w$, $\xi$ and $H$ can be determined by direct measurement of the height correlation function $H(r) = <[h(r)-h(0)]^2>$ with $<...>$ denoting the ensemble average over multiple surface scans [18].

For the sphere roughness we use the measured parameters after 100 nm Au deposition, which is considered bulk as far as optical properties are concerned [4], *w=1.8 nm*, *ξ=22 nm*, *H=0.9 ($\rho_{rms}$=0.23)*. In the following, only the plate roughness was changed. The optical data were obtained from Woollam IR VASE® and VUV-VASE® (Infra red and vacuum ultraviolet variable angle spectroscopic ellipsometer) instruments. (for wavelengths *137nm* to *1.7 μm* and *2 μm* to *33 μm*, respectively) For all calculations on roughness Drude parameters *$w_p$=7.9eV* and *$w_t$=0.048eV are used*. This was obtained by fitting the complex dielectric function in the infrared range of our data. For wavelengths below 137 nm, the data were taken from Palik's handbook.

Figure 1 shows force calculations for a typical film (800 nm thick Au) with *w=7 nm* and *ξ=35 nm*, together with force curves using parameters from hypothetical surfaces with the same *w* but different correlation length *ξ*. Notably the local surface slope for the real surface is *$\rho_{rms}$=0.8* and therefore it is not sufficiently smaller than 1. The inset shows a comparison with the real force data indicating a strong deviation below *50 nm* separation. Therefore, for real films the limits of the perturbation formalism is a serious issue.

The PFA limit is recovered fast with increasing correlation length $\xi$, while differences with the scattering theory are below *5 %* in the range *50-200 nm*. The scattering theory as pointed out in [13] gives the largest deviations in comparison with PFA at large separations. However, in this regime the roughness correction is small (*<1%*). At small separations the PFA becomes more accurate [13], but a comparison with the scattering theory is impossible since the RMS roughness amplitude becomes of the same magnitude as the separation *L*. Therefore, the intermediate separation regime (*~50-200 nm*) is the most interesting range for making a comparison with PFA.

Figure 2 shows the effect of the roughness exponent *H* - limited to relatively high values to avoid large local surface slopes - and arbitrary values for *w* and $\xi$ (indeed $\xi$~*5-50* times the roughness *w*). In this case, deviations from PFA on the force are less than *5%*, but both $\xi$ and *H* have similar effects on the Casimir force. Although for higher local slopes the roughness correction is larger, the latter is not the only cause for this behavior. For this reason we show in the inset of Fig. 2 the difference of the scattering theory ($F_{Scatt}$) and PFA prediction ($F_{PFA}$) for varying roughness amplitudes (since $\sigma(k)$~$w^2$). Thus, we investigate how much increasing *w* enhances the effects predicted by scattering theory compared to the PFA for surfaces with equal local slopes (increase $\xi$ with increasing *w* so that $\rho_{rms}$ is constant). An increased RMS roughness *w* by 4 results in an increase in the difference between scattering and PFA predictions by a factor of only 2. Thus, higher local slopes will give larger effects.

In fact, interesting differences with the PFA start to appear for higher local slopes, where, however, higher order corrections for the scattering perturbation theory are necessary [15]. The observed deviations from the PFA are in the range of claimed experimental accuracy (<5 %) [7-11]. Therefore, the effect of lateral roughness dimensions must be taken into account for high precision measurements or the RMS roughness must be drastically reduced. While it is possible to modify surface roughness (e.g., by annealing and etching etc), the inner structure of the film may be altered as well with such techniques. It was shown in [20] that such effects can be well in the ≤10 % range, and therefore any effect on the Casimir force due to change of surface morphology can be completely offset or even overwhelmed by a different optical response. In Fig. 4 the normalized Casimir force curves are shown for a 100, 200 and 400nm thick film, with respective plasma frequencies 7.8, 6.8 and 6.7 (±0.2) eV and respective relaxation frequencies 48, 40and 38 (±5) meV, obtained by the method mentioned earlier. For comparison a curve for a perfect single crystal film is also shown (well annealed films should give similar response). This curve is obtained by fitting Paliks data and fixing $w_p$ to 9eV (the theoretical value for a single gold crystal).

In conclusion, for a detailed comparison with experimental data, at the level of at least 5 % at short separations (<200 nm), the lateral dimensions of roughness for real films should be included in the theoretical considerations. Moreover, if the RMS roughness is not small, high local surface slopes can have a significant effect (compared to the PFA) on the Casimir force. Our results can be of significance to application related to MEMS/NEMS if Casimir/van der Waals forces are involved and influence the motion of micro-components.

**Acknowledgements:** The research was carried out under project number MC3.05242 in the framework of the Strategic Research programme of the Netherlands Institute for Metals Research (NIMR). Financial support from the NIMR is gratefully acknowledged.

**References**

[1] H. B. G. Casimir, Proc. K. Ned. Akad. Wet. 51, 793 (1948). For initial measurements of the Casimir effect see: M. J. Sparnaay, Physica (Utrecht) 24, 751 (1958); P. H. G. M. van Blockland and J. T. G. Overbeek, J. Chem. Soc. Faraday Trans. 74, 2637 (1978).

[2] K. L. Ekinci and M. L. Roukes, Rev. Sci. Instrum. 76, 061101 (2005)

[3] A. Cleland, Foundations of Nanomechanics (Springer, New York, 2003)

[4] D. Iannuzzi, M. Lisanti, F. Capasso, PNAS 102, 11989 (2005)

[5] F. M. Serry, D. Walliser, and G. J. Maclay, *J. Appl. Phys.* 84, 2501 (1997); [3] Wen-Hui Lin, Ya-Pu Zhao, Chaos, Solitons and Fractals 23, 1777 (2005); G. Palasantzas and J. Th. M. De Hosson, Phys. Rev. B 72, 115426 (2005); G. Palasantzas and J. Th. M. De Hosson, Phys. Rev. B 72, 121409 (2005); G. Palasantzas and J. Th. M. De Hosson, Surf. Sci. 600, 1450 (2006).

[6] R. Onofrio, New J. Phys. 8, 237 (2006)

[7] S. K. Lamoreaux, Phys. Rev. Lett, 78, 5 (1997).

[8] B. W. Harris, F. Chen, U. Mohideen, Phys. Rev. A. 62, 052109 (2000); M. Bordag, U. Mohideen, V. M. Mostepanenko, Phys. Rep. 353 (2001).

[9] H. B. Chan, V. A. Aksyuk, R. N. Kleiman, D. J. Bishop, F. Capasso, Science 291, 1941 (2001)

[10] R. Decca, E. Fischbach, G. L. Klimchitskaya, D. E. Krause, D. Ló´pez, and V. M. Mostepanenko, Phys. Rev. D 68, 116003 (2003).

[11] T. Ederth, Phys. Rev. A, 62, 062104 (2000)

[12] G. Bressi, G. Carugno, R. Onofrio, G. Ruoso, Phys. Rev. Lett. 88 041804 (2002)

[13] P. A. Maia Neto, A. Lambrecht, S. Reynoud, Europhys. Lett. 69 924 (2005); C. Genet, A. Lambrecht, P. Maia Neto and S. Reynaud, Europhys. Lett. 62, 484 (2003).

[14] H. Gies and K. Klingmuller, Phys. Rev. D, 74, 045002 (2006)

[15] P. A. Maia Neto, A. Lambrecht and S. Reynaud, Phys. Rev A 72, 012115 (2005)

[16] P. Meakin Phys. Rep. 235 (1994) 1991; J. Krim and G. Palasantzas, Int. J. of Mod. Phys. B 9, 599 (1995); Y. -P. Zhao, G. -C. Wang, and T. -M. Lu, *Characterization of amorphous and crystalline rough surfaces-principles and applications*, (Experimental Methods in the Physical Science Vol. 37, Academic Press, 2001).

[17] G. Palasantzas, J. Appl. Phys. 97, 126104 (2005)

[18] G. Palasantzas, Phys. Rev. B 48, 14472 (1993); 49, 5785 (1994); G. Palasantzas and J. Krim, Phys. Rev. Lett. 73, 3564 (1994); G. Palasantzas, Phys.Rev.E. 56, 1254 (1997)

[19] Y. -P. Zhao, G. Palasantzas, G. -C. Wang, and J. Th. M. De Hosson, Phys. Rev. B 60, 1216 (1999).

[20] I. Pirozhenko, A. Lambrecht, V. B. Svetovoy, New J. Phys. 8, 238 (2006); For perfect Au films see: E. D. Palik, Handbook of optical constants of Solids (Academic Press, 1995).

**Figure Captions**

**Figure 1** Casimir force for a rough surface $F_{crough}$ divided by the Casimir force $F_c$ for a flat surface expressed in %. Results for different correlation lengths $\xi$ and compared to those of the PFA result (dotted line) for w=7 nm, and H=0.9, and $k_c$=1 nm$^{-1}$. Circles: $\xi$=35 nm (real topography data; $\rho_{rms}$=0.8), Squares: $\xi$=70 nm ($\rho_{rms}$=0.3), Triangles: $\xi$=300 nm ($\rho_{rms}$=0.1). The inset shows the measured force data together with the force calculation from the scattering theory prediction.

**Figure 2** Casimir force for a rough surface $F_{crough}$ divided by the Casimir force $F_c$ for a flat surface expressed in %. Calculations were done with the scattering theory for various roughness exponents $H$ as indicated with w=7 nm, $\xi$=70 nm, and $k_c$=1 nm$^{-1}$. Circles: H =0.7 ($\rho_{rms}$=0.8), Squares: H=0.9 ($\rho_{rms}$=0.3), Triangles: PFA. The inset shows the difference in Casimir Force between scattering theory ($F_{cscatt}$) and PFA ($F_{cPFA}$) divided by that of a flat surface expressed in % for various roughness amplitudes $w$. Circles: *w=14 nm*, Squares: *w=7 nm*, Triangles: *w=3.5 nm*.

**Figure 3** Casimir force for a rough surface $F_{crough}$ divided by the Casimir force $F_c$ for a flat surface expressed in % as a function of the local surface slope $\rho_{rms}$ with w=7 nm, sphere-plate separation 100 nm, and $k_c$=1 nm$^{-1}$. Circles: varying $\xi$ (*24 nm $\leq \xi \leq$ 350 nm*), Squares: varying *H* (*0.48$\leq$H$\leq$0.95*) with *$\xi$=350 nm*.

**Figure 4** Normalized Casimir force for real films with different optical properties, for a 400 nm thick non annealed film (solid line), a 200nm (dotted line), 100nm (circles), and a 'perfect' gold film (triangles).

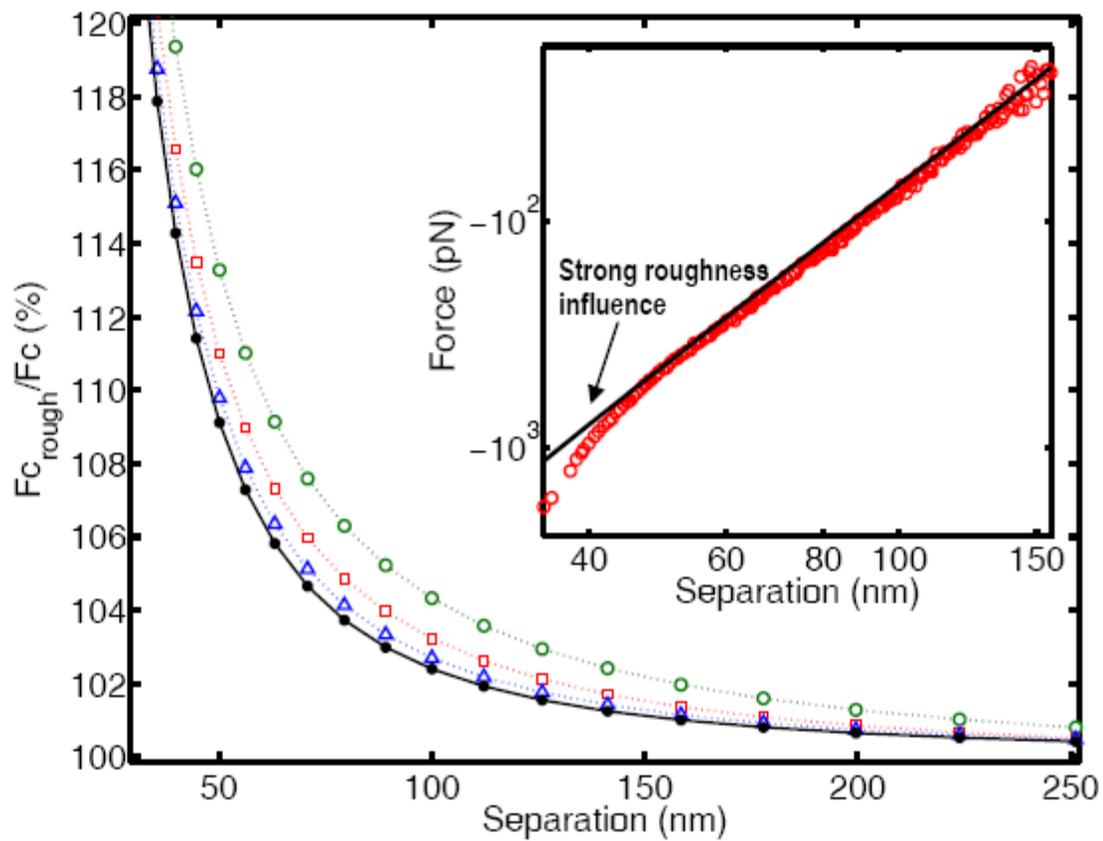

**FIGURE 1**

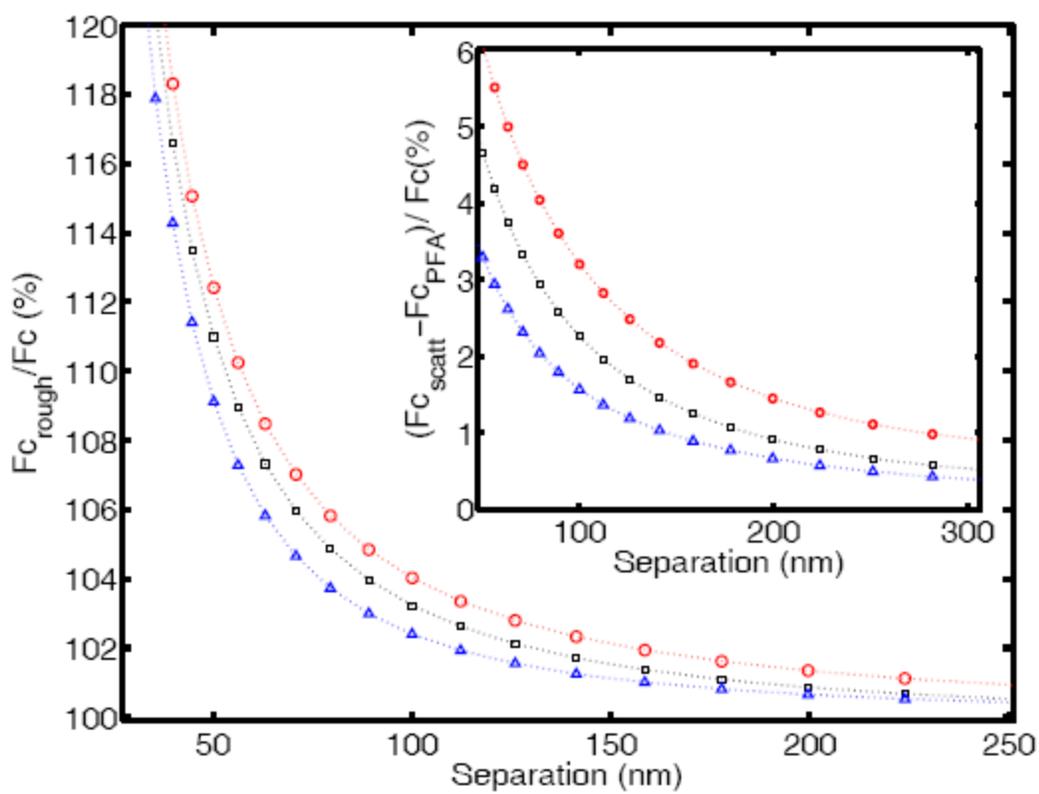

**FIGURE 2**

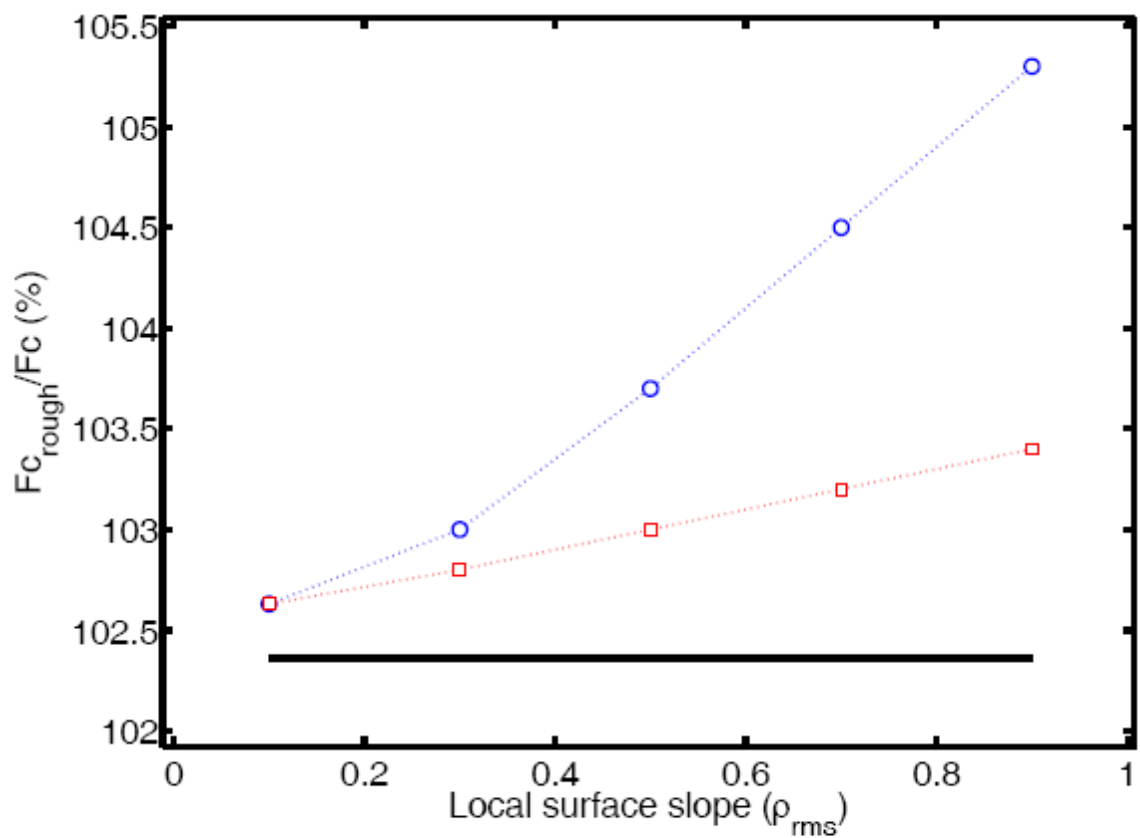

**FIGURE 3**

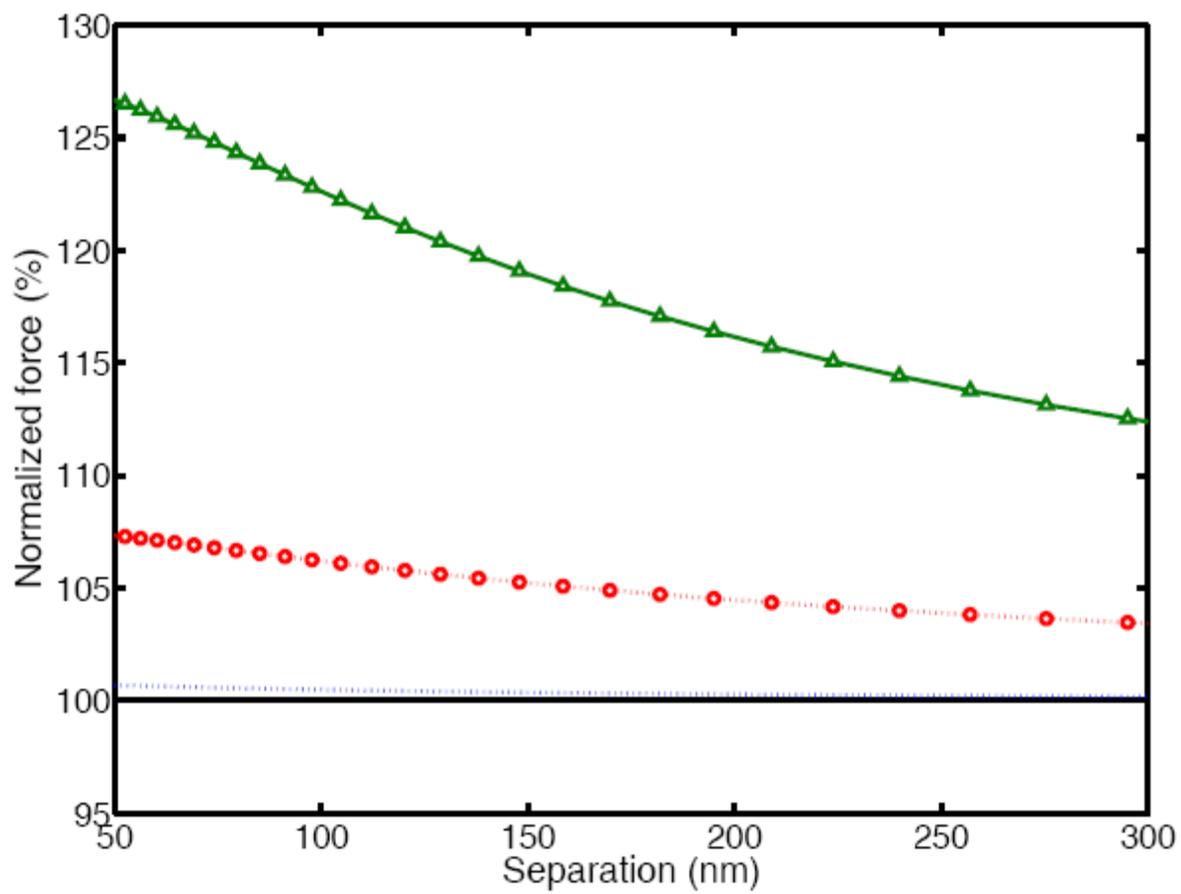

**FIGURE 4**